# Globalization and Glassy Ideality of the Web of Twentieth Century Science

J. C. Phillips

Dept. of Physics and Astronomy, Rutgers University, Piscataway, N. J., 08854-8019

## Abstract

Scientific communication is an essential part of modern science: whereas Archimedes worked alone, Newton (correspondence with Hooke, 1676) acknowledged that "If I have seen a little further, it is by standing on the shoulders of Giants." How is scientific communication reflected in the patterns of citations in scientific papers? How have these patterns changed in the 20$^{th}$ century, as both means of communication and individual transportation changed rapidly, compared to the earlier post-Newton 18$^{th}$ and 19$^{th}$ centuries? Here we discuss a glass model for scientific communications, based on a unique 2009 scientometric study of 25 million papers and 600 million citations that encapsulates the epistemology of modern science. The glass model predicts and explains, using no adjustable parameters, a surprisingly universal internal structure in the development of scientific research, which is essentially constant across the natural sciences, but which because of globalization changed qualitatively around 1960. Globalization corresponds physically to anomalous superdiffusion, which has been observed near the molecular glass transition, and can enhance molecular diffusion by factors as large as 100.

Scale-free network statistics, with power-law exponents, dominate many data bases, including the World Wide Web, the Internet, metabolic networks and organic chemistry [1]. Although power-law scaling with accurate exponents $\gamma$ is observed very widely (37,000+ articles on "power laws", or 4,000+ articles on "power laws" AND scaling* in the Science Citation Index for the last 25 years), at present there exist very few examples where those exponents have been predicted from physical models. Early informetric studies of citation patterns of "only a few" million citations identified such power law patterns [2], which were interpreted in terms of a cumulative advantage model (analogous to properties of equilibrium materials near a critical point; in everyday terms, the rich get richer). Power-law patterns have been interpreted in many interesting ways [1-3]. The exponents represent adjustable parameters that cannot be determined without a specific (preferably physical) model, which should describe systems in equilibrium.

A very large scale scientometric study of $20^{th}$ century citations, involving 25 million papers and 600 million citations [3], unique in the history of epistemology, found that stretched exponentials (SE) give better fits to citation chains for both low and intermediate citation levels $n < n_1$, where $n_1 \sim 40$ earlier in the century, and $n_1 \sim 200$ later. The middle SE region accounts for 95% of the citations, and it, not the scale-free or power law high end, probably objectively represents the essential features of working research citation patterns. The parameters involved in the stretched exponentials appear to follow precise rules already recognized [4] as characterizing ideal glassy relaxation. This coincidence suggests that a universal physical model for systems described by SE distributions can exist, which is analogous to the well-established equilibrium critical point models for power-law distributions, but it must be based on glasses, which are not in equilibrium.

Stretched exponential relaxation (SER) was first recognized by Kohlrausch in 1847 as providing a very good fit (for example, far superior to two exponentials, although the latter involve four parameters) to the residual glassy decay of charge on a Leyden jar, and modern data have increasingly confirmed this superiority, over as many as 12 decades in time for g-Se [4]. The stretched exponential function $A\exp(-(t/\tau)^{\beta})$ contains only three parameters, with $\tau$ being a material-sensitive parameter and $0 < \beta < 1$ behaving as a kinetic fractional exponent. There have been many different theoretical models of SER [4], and for this reason SER is regarded by some as the oldest unsolved problem in science. Broadly speaking, most of these models regard both $\tau$ (material-dependent) and $\beta$ (dimensionless) merely as adjustable parameters, with $\beta$ subject only to weak conditions (such as $1/3 \le \beta \le 1$ [5,6]).

The experimental situation depends on the materials studied. Here we focus on ideal glasses, by which we mean modern microscopically homogeneous glasses [4,7,8], as distinguished from the examples found in older uncritical compilations that included polymer blends, resins and other samples of dubious homogeneity dating back many decades [9]. Included in the "ideal" data base are a few large-scale numerical simulations that explored relaxation over long time periods for non-crystallizing models, for example bidispersive hard spheres. There are nearly 50 of these critically studied ideal examples, spanning a wide range of materials and properties, yet they show a surprisingly simple universal pattern: for glasses dominated by short-range forces only, $\beta = 0.60 \pm 0.01$, while for polymers (long-range intrachain elastic forces) and cases where Coulomb (also long-range) forces were present, $\beta = 0.43 \pm 0.01$. The crossover from 3/5 to 3/7 occurs rarely, but it has been observed in the case of monomer (PG) to polymer (PPG) propyl glycol [7,8].

Fortunately, older theoretical models, including a simple spherical capture model [4,10], had already predicted β = 3/5 before the modern ideal data became available, leaving only the mixed case of short- and long-range forces with β = 3/7 as an unexplained puzzle.  Here an extension of the spherical capture model explains the mixed ideal case in terms of anomalous superdiffusion, which sets the stage for understanding the results of the very large scale scientometric study of 20th century citations, involving 25 million papers and 600 million citations [3].  In this old model [10] excitations diffuse until they are captured by randomly distributed traps.  The result of the diffusion-to-traps model is topological:

$$\beta = d/(d+2) \tag{1}$$

where d is the spatial dimensionality [$(1/3 \leq \beta \leq 1) \leftrightarrow (1 \leq d \leq \infty)$].  Topological relations are independent of geometry, so the model replaces the nearest trap by a spherical trap shell, after which the resulting calculation is straightforward for short-range forces only [10].

Suppose that the diffusive motion is instead determined by competition between short (**r**) - and long-range (**R**) forces, and that the respective length scales are *l* and L.  Also suppose that the average distance between traps is λ (also the radius of the spherical capture construct [10])

$$l < \lambda < L \tag{2}$$

In this intermediate case the diffusive motion becomes entangled, and the simple short-range solution  $\beta_s = d/(d+2)$ is no longer valid.  To see what happens in the entangled case, consider the following example from polymer science.  Dense packing of polymer glasses tends to favor polymer bundles, with individual polymer chains having variable lengths.  Here excitations diffuse to polymer chain ends, which act as traps, but not all the chain ends are at bundle ends; many chains are broken within bundles (length scale λ).  Diffusing excitations can reach these internal traps by diffusing with large steps along adjacent chains (length scale L) bypassing many adjacent traps, or by hopping between chains (short steps, length scale $l$) to reach traps on other chains. This entangled process implies composite relaxation paths that combine small interchain and large intrachain steps.

Heuristically we analyze composite relaxation paths by dividing the excited density $\rho(t)$ harmonically:

$$\rho(t) = \rho_s(t/\tau_s)\rho_l(t/\tau_l) \qquad (3)$$

The local diffusion equation in normal liquids is based on short-range $\mathbf{r}$ forces and is written

$$\partial\rho/\partial t = D\nabla^2_{\mathbf{r}}\,\rho \qquad (4)$$

In glasses with short-range forces the diffusion to traps model leads [10] to (1), $\beta_s = d/(d+2)$, which is in excellent agreement with earlier results [1-4] and those discussed here. How should (4) be generalized to describe the nonlocal case of mixed short $\mathbf{r}$- and long $\mathbf{R}$- range forces? The physical content of (4) is that the spatial fluctuations on an $\mathbf{r}$ scale drive the temporal diffusion. In the mixed $\mathbf{r}:\mathbf{R}$ case, short- (long-)range forces drive spatial fluctuations on an $\mathbf{r}$ ($\mathbf{R}$)scale, with the two processes competing to drive temporal diffusion. This picture is the dynamical relaxation aspect of the "multiple length scale" approach to polarizable static inhomogeneities [11]. Alternatively $\rho$ could be represented as a Fourier sum, and evaluated by an Ewald separation into $\mathbf{r}$ and $\mathbf{R}$ parts. Such nonlocal competition between short- and long-range forces (and small and large (Lévy flight) diffusive steps) suggests the anomalous diffusion equation with superposed fluctuation channels [12]

$$\partial\rho/\partial t = (D_1\nabla^2\mathbf{R} + D_2\nabla^2\mathbf{r})\,\rho \tag{5}$$

Substituting (3) in (5) causes the left hand side to separate into two terms for $\partial\rho_s/\partial t$ and $\partial\rho_l/\partial t$.  The right hand side also separates, and the optimal combination of relaxation by combined short- and long-range forces is determined by the principle of maximum entropy production [13].  One can assume that at the variational extremum this entropic optimization produces a separation of the right hand side of (5) fully parallel to that of the left hand side.  This leads to decoupled equations for $\rho_l(\mathbf{R})$ and $\rho_s(\mathbf{r})$, each equation formally identical to (4).  With $\rho_l(\mathbf{R})$ scaling as $L^{-fd}$, and with $\rho_s(\mathbf{r})$ scaling as $L^{-(1-f)d}$, (5) can then be solved by applying the capture sphere methods of [10] twice to the separately local $\mathbf{r}$ and $\mathbf{R}$ equations to give

$$I(t) \sim \exp[-(t/\tau_1)^{\beta_l}]\,\exp[-(t/\tau_2)^{\beta_s}] \tag{6}$$

with $\beta_l = fd/(fd + 2)$ and $\beta_s = (1-f)d/((1-f)d + 2)$.  The fastest relaxation entropy production [13] at long times will occur along paths for which $f = 1/2$ and $\beta_l = \beta_s = d/(d+4)$, or $3/7$ for $d = 3$.

The extra "push" given to short-range fluctuations by long-range forces explains the anomalously enhanced (by a factor of 100!) surface diffusion reported for fragile molecular ideal glass formers like OTP [14].  Even macromolecules diffusing through small pores exhibit [15] enhanced diffusion, with $\beta \sim 0.8$.  The reduced dimensionality of surfaces or pores enhances the effects of long-range forces.   It is striking that while the measured diffusivity is enhanced by long-range interactions, its long time-dependent relaxation is slowed, as $\beta$ is reduced.

Mathematical note: the physics behind the ideal model describes particles diffusing to traps, whereas the derivation given in [10] utilizes a different geometry, namely the particle is at the sphere center initially and the nearest trap is represented by a spherical shell at $r = \lambda$. However, the final result for $\beta$ is topological $[\beta = \beta(d \text{ only})]$, so the change in geometry is irrelevant. Alternatively, we can consider a random distribution of traps, and partition space into Voronoi polyhedra. We can now construct a topologically isomorphic model, in which the excitations are initially distributed on the polyhedral surfaces. Then we convert this model to its dual, in which the excitations are initially at the polyhedral centers, and the traps are distributed on the surfaces. The dual compartmentalized model preserves d, and yields the same value for $\beta$, which proves the correctness of the spherical cluster construction [10]. It also explains why the bifurcation of $\beta$ into 3/5 and 3/7 channels is so universal in ideal glasses.

The startling result found in the scientometrics study [3] is shown in Fig. 1. Not only do the intermediate citation distributions exhibit SER, but they also shift abruptly in 1960, with stretching exponents $\beta$ that match (within a few %) the $\beta_g = (\beta_{g11}, \beta_{g22}) = (3/5, 3/7)$ values predicted in the 1996 abstract of [4]: 3/5 holds < 1960, while 3/7 holds > 1960! More precisely, the fitted values are $\beta_c = (\beta_{c11}, \beta_{c22}) = (0.57, 0.47)$. The task of theory, to explain these results, goes well beyond noting their agreement with the d = 3 glass values identified in 1996, as one must justify using d = 3, and also explain how the microscopically homogeneous conditions necessary for glass networks become appropriate for the world-wide 20$^{\text{th}}$ century citation network. One must also explain how these conditions changed abruptly in 1960, and here this is done without using adjustable parameters. Finally

$$\beta_{g11} + \beta_{g22} = \beta_{c11} + \beta_{c22} \qquad (7)$$

to within 1%.

The quantitative correspondences between citation network distributions and glass network relaxation are precise, and already break down for even deeply supercooled liquids, with viscosities millions of times that of water. The characteristic features of glass networks are that the atoms are confined to cages, within which they vibrate for long times without escaping [16]. The cages are formed because there is a near equality between the number of short-range force constraints and the number of spatial degrees of freedom d; the usual momentum degrees of freedom, also d and found in gases and liquids, are absent in glasses.

When an article is written, it utilizes the existing literature as sources of its citations, so the cited articles correspond to the caged atoms in a glass. The citing article corresponds to an excited atom of the glass, which as it diffuses through the glass, undergoes collisions with caged atoms, each collision representing a citation. When the excited atom stops, the citation chain ends, and the last citation corresponds to a fixed (immobile) trap. The trap captures the excited atom, which becomes immobile and part of the glass network. The correspondence between citation statistics and glass relaxation occurs because the latter is also described most accurately by a diffusion-to-immobile traps model. In supercooled liquids the traps are mobile; their sweeping motion accelerates their capture of excitations, shortening lifetimes and increasing the stretching fraction β above its glass value, and destroying the correspondence with the citation stretching fractions.

The apparent citation dimension is likely to be d = 3 (latitude, longitude, and time $t$). Since there is a strong tendency to cite articles that one knows best, and these are often the articles of one's closest colleagues, and possibly their closest colleagues, the citation number for a given article will be limited largely to this circle. This corresponds to $\beta_1 = 3/5 \sim 0.57$, which gives the best fit to low and intermediate citation levels prior to 1960.

Because the detailed aspects of a research article are best known only to this small space-time circle, only the more general aspects will give rise to a larger number of citations. After 1960, scientific conferences became popular and grew steadily in size, for a wide variety of practical reasons – an upsurge in funding for science after Sputnik (suddenly many research scientists were offered all-expenses paid intercontinental travel), rapid moderation of commercial jet fares (the first broadly successful commercial jet, the Boeing 707, was delivered in waves, with the first delivery wave peaking in 1960), etc. At international conferences one encounters a wider range of ideas in a much less detailed format, which corresponds to long-range interactions, and appears to give a broader distribution of citations. The wider range of ideas can survive longer, the article can continue to accumulate citations, and because $f \sim 1/2$ the diffusive interactions affect only details, while leaving the larger ideas intact and still capable of accumulating citations. Here it is quite striking that the balance between short- and long-range effects on relaxation in glasses should be so closely echoed (similar values of f!) in the competition between narrower and wider influences on citation patterns.

If we regard $\beta_c$ and $\beta_g$ as the diagonal elements of $\beta_{ij}$ matrices (i,j =1,2) in citation and glass spaces respectively, then (7) is explained by conservation of the traces of the $\beta_c$ and $\beta_g$ matrices under unitary transformations, and we do not need to know any details about the local and mixed forces interactions. To explain the reduction in the splitting of the informetric citation values relative to physical values found in glasses, we must assume that while the glass matrix can be taken as diagonal except in rare crossover cases [7], the citation matrix is anti-Hermitian, with identical imaginary off-diagonal values $\beta_{c12} = \beta_{c21}$ = ai, with a << 1. Some recent physical examples [17] of anti-Hermitian matrices are discussed elsewhere [8].

There are many precise ideal physical examples [4,7,8] of competing short- and long-range forces. Elegant studies of decay of luminescence from 16 single-crystal isoelectronic $ZnSe_{1-x}Te_x$ alloys of commercial quality (these microscopically homogeneous alloys are used in orange light-emitting diodes) [18] exhibited SER with a fitting accuracy of 0.2%. Two previously unnoticed extrema in $\tau(x)$ or $\beta(x)$ are associated with $\beta$ = 3/5 or 3/7 [8]. Simulations of Josephson junction arrays gave $\beta$ = 0.45 over a wide range $4/9 < T/T_c < 2/3$ [19]. The long-range interactions involve release or capture of vortex-anti-vortex pairs by nanodomain walls (length L), as illustrated in their Fig. 3. The arrays represent an excellent realization of the geometry assumed in [10], but with long L hops included. The checkerboard patterns observed in their Fig. 2 are similar to those reported for rigidity percolation in network glasses [20].

## Figure Caption

Fig. 1. Evolution of the number of articles n with a number of citations τ (or T) as a function of the decade in which the article was published. The distributions for the decades up to 1960 bunch and are well fitted by an SE with $\beta_1 = 0.57$, while the decades after 1960 also bunch, but are fitted with $\beta_2 = 0.47$. There is no crossover decade; the abrupt crossover at 1960 is unambiguous [3], as are the accuracies of the SE fits.

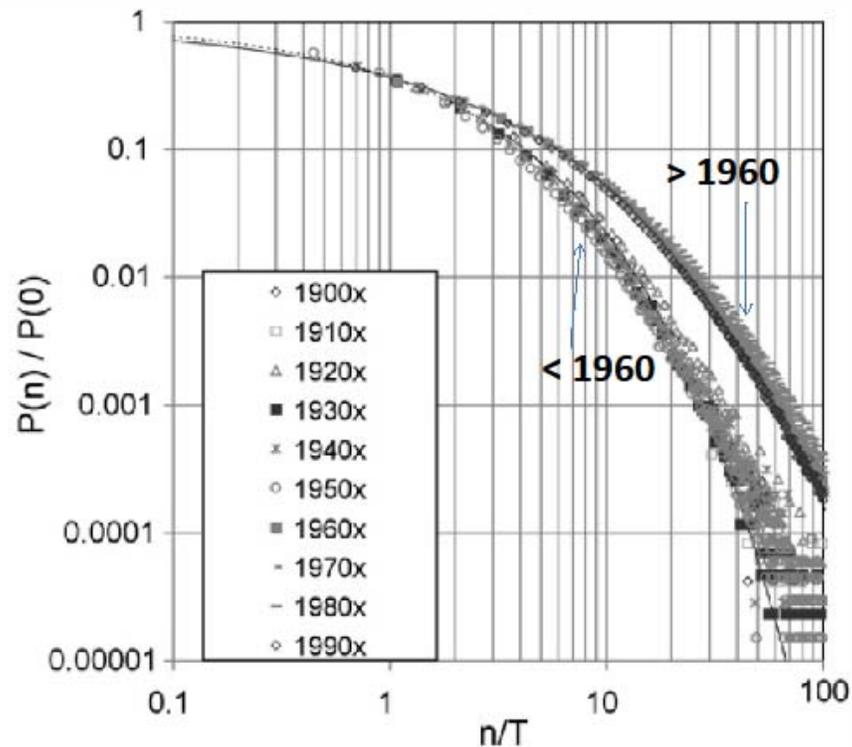

Fig. 1.